\hfuzz 2pt
\vbadness 10000
\font\titlefont=cmbx10 scaled\magstep1

\magnification=\magstep1

\null
\vskip 1.5cm
\centerline{\titlefont ENTANGLEMENT GENERATION}
\medskip
\centerline{\titlefont IN THE UNRUH EFFECT}
\vskip 2.5cm
\centerline{\bf F. Benatti}
\smallskip
\centerline{Dipartimento di Fisica Teorica, Universit\`a di Trieste}
\centerline{Strada Costiera 11, 34014 Trieste, Italy}
\centerline{and}
\centerline{Istituto Nazionale di Fisica Nucleare, Sezione di 
Trieste}
\vskip 1cm
\centerline{\bf R. Floreanini}
\smallskip
\centerline{Istituto Nazionale di Fisica Nucleare, Sezione di 
Trieste}
\centerline{Dipartimento di Fisica Teorica, Universit\`a di Trieste}
\centerline{Strada Costiera 11, 34014 Trieste, Italy}
\vskip 2cm
\centerline{\bf Abstract}
\smallskip
\midinsert
\narrower\narrower\noindent
The master equation describing the completely positive time evolution
of a uniformly accelerated two-level system in weak interaction
with a scalar field in the Minkowski vacuum is derived and explicitly
solved. It encodes the well known Unruh effect, leading to a purely
thermal equilibrium state. This asymptotic state turns out to be entangled
when the uniformly accelerating system is composed by two,
independent two-level atoms.
\endinsert

\vfill\eject

{\bf 1. INTRODUCTION}
\bigskip
 
When a particle detector moves with a uniform
acceleration through an external field in its vacuum state, 
spontaneous excitation can occur; indeed, the detector behaves as if it were
in a thermal bath, with temperature (the ``Unruh temperature'')
proportional to its proper acceleration. 
This phenomenon is known as the ``Unruh effect''.[1-6]
Although originally presented for detectors in interaction with massless, free
scalar fields, the effect has been later confirmed in the case of massive and
higher spin fields, in arbitrary spacetime dimensions (with appropriate
gray-body corrections to the thermal bath spectrum),[5] and even for arbitrary
interacting fields, using the algebraic formulation of quantum field theory.[7, 8]
Experimental verification of the phenomena is hard to be realized in practice,
since measurable thermal effects require very high accelerations; nevertheless,
clever proposals involving circular trajectories have been discussed
and might get realized in the future.[9]

Most treatments of the Unruh effect focus on the study of the spontaneous
excitation of an accelerated ``DeWitt detector'',[10] a non-relativistic
$n$-level system, linearly coupled to the external relativistic field.
What is then studied is the excitation rate, {\it i.e.} the probability
per unit time of a spontaneous transition from the ground state
to one of its excited states.
This probability clearly vanish for a detector at rest, but turns out
to be non-zero, and proportional to a Planckian factor, for a uniformly
accelerated one. The same physical result can be obtained by examining
the vacuum state of the external field as seen by the accelerating detector:
it turns out to be thermal, and described by the same Planckian factor.

This result, known as the ``thermalization theorem'',[5] implies that the
accelerating detector behaves like an open system, {\it i.e.}
a system immersed in an external heat bath.\hbox{[11-15]} In this respect,
the study of its full dynamics and not only of its asymptotic
excitation rate is of great physical interest. As for any open system,
it can be obtained from the complete time-evolution describing the
total system (detector + external field) by integrating over the field
degrees of freedom, that in fact are never observed.

In the following, we shall explicitly derive and study such reduced dynamics,
restricting our attention to the case of a two-level detector in interaction
with free, massless, scalar fields. As in the standard analysis
of the Unruh effect, this simplified situation is not really restrictive
since it is able to capture all the main features of the thermalization phenomena,
without the algebraic complications needed to treat the case of higher-spin or
massive fields.

In addition, we shall assume the interaction between the accelerating two-level
system and the scalar fields to be weak;[12-14] while encompassing the most common physical
situations, this hypothesis allows a rigorous, mathematically sound, derivation
of the evolution equation describing the dynamics of the moving detector.[16, 17]
The corresponding finite-time evolution takes the form of a one-parameter semigroup
of completely positive maps,[11-14] whose action on the state of the system can be
explicitly obtained in closed form. Explicit expression for the probability 
transitions among the states of the moving detector can then be obtained:
they contain appropriate Planckian factors, so that the known forms of the
spontaneous excitations rates are reproduced. Indeed, one finds that the
equilibrium configuration of the detector is exactly thermal, with temperature
equal to the Unruh temperature. This is just another manifestation of
the ``thermalization theorem'': a single, two-level system accelerating through
a vacuum field is asymptotically driven to a thermal equilibrium with the
bath.

A physically much more interesting asymptotic situation can be obtained when 
the uniformly accelerating system is formed by two, independent two-level
systems, in weak interaction with the same scalar field.
The corresponding master equation that describes the reduced dynamics of
the two detectors can be explicitly written generalizing the results obtained
in the case of a single detector. It again generates a semigroup of completely
positive maps, whose asymptotic equilibrium state can be easily determined.
In general, it turns out to be an entangled state, even in the case of
a separable initial state.

It is known that, in certain circumstances, heat baths belonging to a specific class
can enhance entanglement rather than destroying it:[18-23] two, mutually non-interacting 
systems immersed in one of these baths can then become quantum correlated; 
this can happen both at finite time and in the asymptotic regime.
It is remarkable that the thermal bath seen by uniformly accelerating systems
precisely belongs to the mentioned class.
In turn, this result may suggest a new possibility for an experimental
test of the Unruh effect: use appropriate quantum optics devices to detect the
asymptotic entanglement generated by the uniform acceleration.

The next Section will be devoted to the discussion of the weak
coupling limit and the derivation of the master equation describing
the dynamics of the accelerating system in its comoving frame.
The finite time evolution it generates will be analyzed in detail
in Section 3 for a single two-level system.
The case of a detector composed by two, independent two-level systems
will be instead treated in Section 4, and the possibility of
generation of mutual quantum correlations analyzed; in particular,
the entanglement content of the asymptotic equilibrium state 
will be determined by computing its corresponding concurrence.
Finally, Section 5 contains further physical considerations
on the open system approach to the Unruh effect,
while the Appendix illustrates some technical issue 
connected with the derivation of the master equation.

\vskip 2cm

{\bf 2. MASTER EQUATION}
\bigskip

As explained in the introductory remarks, we shall use well-known techniques 
developed in the study of open quantum systems to analyze the behaviour of 
an idealized detector immersed in a scalar vacuum field, moving along
a prescribed, uniformly accelerating trajectory.

A frame-independent description of the dynamics of such a system would
require a covariant formulation of the evolution equation of quantum mechanics.%
\footnote{$^\dagger$}{Combining quantum mecahnics and relativity in a fully
consistent way is notoriously problematic; for recent reviews on different
aspects of this question, see Refs.[24, 25].}
Although rather cumbersome in practice, such a description exists: 
it has been introduced in the early days of quantum field theory,
and it is based on the so-called Schwinger-Tomonaga equation.[26] It extends
the standard, hamiltonian evolution to a local functional dynamics,
where the time variable is replaced by a
collection of three-dimensional spacelike hypersurfaces,
upon which spacetime has been sliced.[15, 27]

As an alternative to this rather complicated approach,
one can fix {\it ab initio} a reference frame and study the dynamics
of the system with respect to the natural (albeit in general only locally valid)
time variable there defined. 
This choice is particularly convenient when a global time-variable is
available, as in an inertial or in an uniformly accelerated frame,[3] since
it allows drastic simplifications in the analysis and interpretation of
the system dynamics.
The price to pay is the lack of manifest covariance and thus of a simple way of
translating a physical description obtained in one reference frame to that
derived in a different one.

Both approaches are in line of principle always available,
although, for practical reasons, one usually 
works in the ``laboratory'' frame. 
However, in certain physical situations, 
the use of the covariant formulation may in practice be precluded.
This is indeed the case of a system in interaction with an external
environment: in fact, a consistent, reduced dynamics can be derived only 
within reference frames for which environment correlations decay 
sufficiently fast.[12-15] Notice that, in most cases, this is not really a serious drawback,
since these preferred reference frames are precisely the ones used
in the experiments.

For the system under study, since one is interested in the behaviour of the accelerating
detector, the natural reference frame to be adopted is that comoving with it:
its dynamics will then be described as an evolution in the proper time $t$.
In this reference frame, the detector is always at rest, and without loss of
generality, it can be positioned at the origin of the comoving spatial coordinates.

In the inertial reference frame, with Minkowski coordinates $(x^0,x^1,x^2,x^3)$,
the detector is instead seen following an hyperbolic trajectory.
More specifically, in order to better conform with the standard open system
paradigm adopted below, we shall assume that the composed system
$\{$detector + external fields$\}$ be initially prepared in a factorized state,
with the detector at rest and the fields in their vacuum state.
The detector starts moving at time $t\equiv x^0=\,0$, after which
it follows the path:
$$
\eqalign{
&x^0(t)={1\over a}\sinh at\ ,\cr
&x^1(t)={1\over a}\cosh at\ ,\cr
&x^2(t)=x^3(t)=\, 0\ ,
}
\eqno(2.1)
$$
with $a$ the constant proper acceleration.%
\footnote{$^\dagger$}{This situation looks more physical than the usually adopted
one for which the detector follows the hyperbolic motion for all times.}
Our aim is to derive and study the 
appropriate reduced ``master equation'' that describes the quantum evolution
of the detector following this trajectory for positive $t$.

To this aim, the details of the detector internal dynamics result
irrelevant: one can then choose to model it as a simple two-level atom,
{\it i.e.} as a non-relativistic, quantum mechanical system,
that can be fully described in terms of a two-dimensional
Hilbert space. With respect to a fixed, arbitrary basis in this space,
states of the system will be represented
by a $2\times 2$ density matrix $\rho$, {\it i.e.} an hermitian,
$\rho^\dagger=\rho$, normalized, ${\rm Tr}[\rho]=1$,
operator, with non-negative eigenvalues, ${\rm det}[\rho]\geq0$.

In absence of any interaction with the external scalar fields,
the atom internal dynamics will be driven by a $2\times 2$
hamiltonian matrix $H_S$, that in the chosen basis can be taken
to assume the most general form:
$$
H_S={\omega\over 2}\sum_{i=1}^3 n_i\sigma_i\equiv
{\omega\over 2}\, \vec n\cdot\vec\sigma\ ,
\eqno(2.2)
$$
where $\sigma_i$, $i=1,2,3$ are the Pauli matrices, $n_i$, $i=1,2,3$
are the components of a unit vector, while $\omega$ represents the
gap between the two energy eigenvalues. 

As mentioned in the Introduction, the interaction of the atom 
with the external scalar fields is assumed to be weak; it can then
be described by an hamiltonian $H'$ that is linear in both atom and field
variables:
$$
H'=\sum_{\mu=0}^3\sigma_\mu\, \Phi_\mu\big(x(t)\big)\ ,
\eqno(2.3)
$$
with $\sigma_0$ the $2\times 2$ unit matrix. 
As explicitly indicated, the interaction is
effective only along the trajectory (2.1), as the
atom is assumed to be an idealized point, without size.%
\footnote{$^\dagger$}{For a detailed discussion on the physical motivations
justifying this simplifying assumption, see Ref.[5].}
The operators (or better, operator valued distributions) $\Phi_\mu(x)$
represent the external fields and satisfy the massless Klein-Gordon equation.
They can be expanded as:
$$
\Phi_\mu(x)=\sum_{a=1}^N \Big[ \chi_\mu^a\, \phi_a^{(-)}(x)
+\big(\chi_\mu^a\big)^*\, \phi_a^{(+)}(x)\Big]\ ,
\eqno(2.4)
$$
in terms of positive- $\phi_a^{(+)}(x)$ and negative-energy
$\phi_a^{(-)}(x)$ field operators relative to a set of $N$ independent, massless,
free scalar fields, with total hamiltonian $H_\Phi$; the complex coefficients
$\chi_\mu^a$ ``embed'' the field modes into the two-dimensional
detector Hilbert space and play the role of (generalized) coupling constants.
The explicit form of the field hamiltonian $H_\Phi$ need not be specified;
it suffices to know that once transformed in the inertial reference frame
it leads to the standard expansion of $\phi_a^{(+)}$, $\phi_a^{(-)}$
in terms of Minkowski creation, annihilation operators, respectively.

In the comoving frame, the total hamiltonian $H$ for the
complete system $\{$atom + external fields$\}$ can thus be written as
$$
H=H_S + H' + H_\Phi\ .
\eqno(2.5)
$$
It generates the evolution in $t$ of the corresponding 
total density matrix $\rho_{\rm tot}$,
$$
{\partial\rho_{\rm tot}(t)\over\partial t}=-i\, L_H[\rho_{\rm tot}(t)]\ ,
\eqno(2.6)
$$
starting at $t=\,0$ from the initial
configuration: $\rho_{\rm tot}(0)=\rho(0)\otimes |0\rangle\langle 0|$,
where $|0\rangle$ is the Minkowski field vacuum state, as seen in the comoving frame;
the symbol $L_H$ in (2.6) represents the Liouville operator corresponding to $H$,
$$
L_H[\,\cdot\,]\equiv[H, \ \cdot\ ]\ .
\eqno(2.7)
$$
The dynamics of the atom is then obtained by summing over 
the field $\Phi$ degrees of freedom, 
{\it i.e.} by applying to $\rho_{\rm tot}(t)$ the trace projection operator $P$:
$$
\rho(t)=P\big[\rho_{\rm tot}(t)\big]\equiv 
{\rm Tr}_\Phi \big[\rho_{\rm tot}(t)\big]={\rm Tr}_\Phi\Big[ e^{-itH}\, 
\Big(\rho(0)\otimes |0\rangle\langle 0|\Big)\, e^{itH}\Big]\ ;
\eqno(2.8)
$$
it is generated by an equation that is the result
of the action of $P$, and its complement operator $Q=1-P$, on both sides
of (2.6). It can be conveniently written as an integro-differential equation:[12]
$$
{\partial\rho(t)\over\partial t}=-i\, L_{H_S}[\rho(t)]+\int_0^t ds\, {\cal K}_s[\rho(t-s)]\ ,
\eqno(2.9)
$$
where the time-dependent kernel ${\cal K}_s$ is defined as a formal power 
series expansion in the interaction hamiltonian $H'$,
and it is explicitly given by the expression:
$$
{\cal K}_s[\rho]=-\,{\rm Tr}_\Phi\Big\{L_{H'}\, Q\, e^{-is Q L_H Q}\, Q\, 
L_{H'}\big[\rho\otimes |0\rangle\langle 0|\big]\Big\} \ .
\eqno(2.10)
$$

The resulting finite evolution map $\rho(0)\mapsto \rho(t)$
is in general very complicated, developing irreversibility and memory effects;
indeed, the atom state $\rho(t)$ at time $t$ as given by (2.9) 
depends not only on the initial state
$\rho(0)$, but also on all states $\rho(s)$, with $s<t$. This is a general result,
valid for any reduced dynamics: nevertheless, the form of the master equation (2.9)
can be further simplified on the basis of additional physical considerations,
that in the present case amounts to the requirement of a weak coupling
between the moving atom and the external fields.[11-15]

On general grounds, one expects the memory effects in (2.9) to be negligible
when the ratio $\tau/\tau_\Phi$ between the typical variation time $\tau$
of $\rho(t)$ and the decay time $\tau_\Phi$ of the field correlations is large.
This is precisely the result of the ``weak coupling limit'' procedure:[13] it first
amounts to a rescaling of the interaction hamiltonian $H'$ by a dimensionless
coupling constant $g$, so that the first contribution
in the r.h.s. of equation (2.9), representing the unperturbed motion of the atom, 
is of order one,
while the remaining piece, taking care of the interaction with the external field,
becomes of order $g^2$.
This implies that the evolution of the state $\rho(t)$ develops 
on time scales of order $1/g^2$; in order to obtain a consistent
physical description, one then needs to appropriately rescale the time
variable, $t\to t/g^2$. After this, $g$ can be safely taken to be vanishingly small,
and the evolution equation attains a well defined limit.

The procedure that we have briefly outlined
can be given a precise mathematical meaning:[16, 17] provided the field correlations
decay sufficiently fast at large time separations, one can rigorously prove that
in the weak coupling limit the formal expression in (2.9), 
(2.10) converges to a differential equation for the reduced density
matrix $\rho(t)$, that is local in time. It takes the so-called
Kossakowski-Lindblad form (see the Appendix for details):[28, 29]
$$
{\partial\rho(t)\over \partial t}= -i \big[H_{\rm eff},\, \rho(t)\big]
 + {\cal L}[\rho(t)]\ ,
\eqno(2.11)
$$
with
$$
{\cal L}[\rho]={1\over2} \sum_{i,j=1}^3 a_{ij}\big[2\, \sigma_j\rho\,\sigma_i 
-\sigma_i\sigma_j\, \rho -\rho\,\sigma_i\sigma_j\big]\ .
\eqno(2.12)
$$
The corresponding finite-time evolution maps $\gamma_t$
generated by this equation, $\rho(0)\mapsto \rho(t)=\gamma_t[\rho(0)]$,
form a one-parameter semigroup of trace-preserving transformations:
$\gamma_t\circ\gamma_s=\gamma_{t+s}$, $t,s\geq0$.

The effective hamiltonian $H_{\rm eff}$ and the coefficients of the $3\times 3$ 
Kossakowski matrix
$a_{ij}$ depend on the Fourier transform of the field vacuum correlations
(Wightman functions), evaluated along the trajectory (2.1):
$$
\alpha_{\mu\nu}(\lambda)=\int_{-\infty}^\infty dt\ e^{i\lambda t}\
\langle 0| \Phi_\mu\big(x(t)\big)\, \Phi_\nu(0) |0\rangle\ .
\eqno(2.13)
$$
Being invariant,
these correlation functions can be computed in any reference frame. Recalling that
the field variables $\phi_a^{(\pm)}$ in (2.4) are all independent,
one finds:
$$
\langle 0| \Phi_\mu(x)\, \Phi_\nu(y) |0\rangle= \sum_{a=1}^N \chi^a_\mu \big(\chi_\nu^a\big)^*\
G(x-y)\ ,
\eqno(2.14)
$$
where $G(x-y)$ is the standard four-dimensional Wightman function for a single scalar field,
that, with the proper $i\varepsilon$ prescription, can be written as%
\footnote{$^\dagger$}{The existence of the weak-coupling limit is guaranteed by the
convergence of the integral $\int_0^\infty dt\ |G(x(t))|\, (1+t)^\delta$, for some
$\delta>0$ (for details, see Ref.[16]); this is assured at infinity by the exponential fall off of
$G(x(t))$, and at zero by the $i\varepsilon$ prescription.[5]}
$$
G(x)=\int {d^4 k\over (2\pi)^3}\, \theta(k^0)\,\delta(k^2)\, e^{i k\cdot x-\varepsilon k^0}
=-{1\over 4\pi^2}{1\over (x^0-i\varepsilon)^2 - (\vec x)^2}\ .
\eqno(2.15)
$$
Its Fourier transform along the trajectory (2.1) can be easily evaluated through
a contour integral:[5]
$$
{\cal G}(\lambda)=\int_{-\infty}^\infty dt\ e^{i\lambda t}\ G\big(x(t)\big)=
{1\over 2\pi}\, {\lambda\over 1-e^{-\beta_U\lambda}}\ ,
\eqno(2.16)
$$
where $\beta_U=1/T_U$, with $T_U=a/2\pi$, the so-called Unruh temperature.
With the help of the $3\times 3$ hermitian matrices (see Appendix),
$$
\psi_{ij}^{(0)}=n_i\, n_j\ ,\qquad 
\psi_{ij}^{(\pm)}={1\over 2}\big(\delta_{ij} - n_i\, n_j\pm i\epsilon_{ijk} n_k\big)\ ,
\eqno(2.17)
$$
it proves convenient to define the transformed coupling coefficients
$\chi_i^{(\xi)}{}^a=\sum_j\chi^a_j\, \psi_{ji}^{(\xi)}$,
with $\xi=0,+,-$,
together with their corresponding complex conjugate ones
$\overline\chi_i^{(\xi)}{}^a=\sum_j\big(\chi^a_j\big)^*\, \psi_{ji}^{(-\xi)}$;
by means of them, the Kossakowski matrix $a_{ij}$ in (2.12) explicitly reads:
$$
a_{ij}=\sum_{a=1}^N \Big[{\cal G}(0)\, \chi_i^{(0)}{}^a\, \overline\chi_j^{(0)}{}^a
+{\cal G}(\omega)\, \chi_i^{(+)}{}^a\, \overline\chi_j^{(+)}{}^a
+{\cal G}(-\omega)\, \chi_i^{(-)}{}^a\, \overline\chi_j^{(-)}{}^a\Big]\ .
\eqno(2.18)
$$
Being the sum of three manifestly positive terms, the hermitian matrix
$a_{ij}$ turns out to be positive. As a consequence, the one parameter
family of transformations $\gamma_t$ generated by the equation (2.11), (2.12) is composed
by completely positive maps.[12-14, 30] As well known, this property
assures the positivity of the evolved density matrix
$\rho(t)$ in any physical situation, thus guaranteeing the correct interpretation
of its eigenvalues as probabilities. (For discussions on this relevant
point, see Refs.[31-34].)

This result is non-trivial and shows the importance of adopting
a physically consistent and mathematically precise procedure, 
the Davies weak coupling limit, 
in deriving the reduced evolution equation.[35] Indeed, note that
direct use of the standard second order perturbative approximation
in the original master equation (2.9) (as adopted in [36]) 
produces a finite time evolution
for $\rho(t)$ that in general does not preserve the positivity
of probabilities.

Besides producing the non-unitary evolution term (2.12), the coupling with
the external fields $\Phi_\mu$ induces also a correction to the system hamiltonian (2.2), 
the so-called Lamb shift $H_L$: the
complete hamiltonian is now $H_{\rm eff}=H_S+H_L$. As for the Kossakowski matrix,
this additional shift $H_L$ can be expressed in terms of the field correlations
along the accelerating trajectory.
Introducing together with the Fourier transforms (2.13) 
also their corresponding Hilbert transforms:[12, 14]
$$
\beta_{\mu\nu}(\lambda)={P\over i\pi}\int_{-\infty}^\infty dz\ {\alpha_{\mu\nu}(z)\over
z-\lambda}\ ,
\eqno(2.19)
$$
where $P$ denotes principal value, one finds:
$$
H_L={1\over2}\sum_{i=1}^3 b_i\,\sigma_i\ ,
\eqno(2.20)
$$
with
$$
\eqalign{
b_i=&\,i\sum_{j=1}^3\Big[\alpha_{0j}(0)-\alpha_{j0}(0)-
\beta_{0j}(0)-\beta_{j0}(0)\Big]\,\psi_{ji}^{(0)}\cr
+&\sum_{k,l,r,s=1}^3\epsilon_{ikl}\, \Big[\beta_{rs}(0)\,\psi_{rk}^{(0)}\psi_{sl}^{(0)}
+\beta_{rs}(\omega)\,\psi_{rk}^{(+)}\psi_{sl}^{(-)}
+\beta_{rs}(-\omega)\,\psi_{rk}^{(-)}\psi_{sl}^{(+)}\Big]\ .
}
\eqno(2.21)
$$
This expression is however formal and requires renormalization. 
It involves the following integral transform
of the scalar Wightman function (compare with (2.19)),
$$
{\cal K}(\lambda)={P\over i\pi}\int_{-\infty}^\infty dz\ 
{{\cal G}(z)\over z-\lambda}\ ,
\eqno{(2.22)}
$$
which, recalling (2.16), can be split as
$$
{\cal K}(\lambda)={P\over 2\pi^2 i}\int_0^\infty dz\ {z\over z-\lambda}+
{P\over 2\pi^2 i}\int_0^\infty dz\ {z\over 1-e^{\beta_U z}}
\bigg[{1\over z+\lambda}-{1\over z-\lambda}\bigg]\ ,
\eqno(2.23)
$$
into an inertial and an acceleration dependent piece. 
Although not expressible in terms of elementary functions,[37] the acceleration
dependent second term is a finite, odd function of $\lambda$, vanishing
as $\beta_U$ becomes large, {\it i.e.} for a vanishing acceleration.
The first contribution in (2.23) is however linearly divergent.
As a consequence, despite some cancellations 
that occur in (2.21) (see below), the Lamb contribution $H_L$ turns
out to be infinite, and its definition requires the introduction 
of a suitable cutoff and a renormalization procedure.

This is a well known fact, and has nothing to do with the
weak-coupling assumptions used in deriving the evolution equation (2.11), nor
with the specific situation of an accelerating atom: as the splitting 
in (2.23) shows, the Lamb shift would be infinite 
even for an atom at rest. Rather, the appearance of the
divergences is due to the non-relativistic treatment of
the moving two-level atom, while any sensible calculation
of energy shifts would have required the use 
of quantum field theory techniques.[38, 39]

In our quantum mechanical setting, the procedure needed
to make the Lamb contribution $H_L$ well defined is therefore
clear: perform a suitable
acceleration independent subtraction, so that
the expression in (2.21) reproduces the correct
quantum field theory result when the atom is at rest.
However, since we are interested in analyzing the effects 
due to the uniformly accelerated motion of the atom,
we do not need to do this explicitly. 
In the following we shall therefore ignore standard, acceleration-independent
hamiltonian contributions in the evolution equation (2.11)
and concentrate on the phenomena induced by the motion of the atom.

\vskip 2cm

{\bf 3. SINGLE ACCELERATING ATOM AND DECOHERENCE}
\medskip

We shall now  explicitly discuss the dynamics of the accelerated
atom as described by the evolution equation (2.11), (2.12).
In order to simplify a bit the treatment and be able to write
explicit expressions for the evolved state $\rho(t)$,
we shall assume that the coupling coefficients $\chi^a_\mu$
introduced in (2.4) satisfy the further condition:
$$
\sum_{a=1}^N \chi^a_\mu \big(\chi_\nu^a\big)^*\propto\delta_{\mu\nu}\ :
\eqno(3.1)
$$
in this way, the field correlations in (2.14) become diagonal;
in the following, for sake of simplicity, we shall set to one the
proportionality coupling coefficient.
Then, the sum in (2.18) can be explicitly performed
and the Kossakowski matrix $a_{ij}$ takes the general form
$$
a_{ij}=A\, \delta_{ij}-iB\, \epsilon_{ijk}\, n_k + C\, n_i\, n_j\ ,
\eqno(3.2)
$$
where
$$
\eqalign{
&A={1\over2}\Big[{\cal G}(\omega)+{\cal G}(-\omega)\Big]={\omega\over 4\pi}
\bigg[{1+e^{-\beta_U\omega}\over 1-e^{-\beta_U\omega}}\bigg]\ ,\cr
&B={1\over2}\Big[{\cal G}(\omega)-{\cal G}(-\omega)\Big]={\omega\over 4\pi}\ ,\cr
&C={1\over2}\Big[2{\cal G}(0)-{\cal G}(\omega)-{\cal G}(-\omega)\Big]={\omega\over 4\pi}
\bigg[{2\over\beta_U\omega}-{1+e^{-\beta_U\omega}\over 1-e^{-\beta_U\omega}}\bigg]\ .\cr
}
\eqno(3.3)
$$
Similarly, also the Lamb shift contribution $b_i$ in (2.21) simplifies, and become
directed along the unit vector $\vec n$; the effective hamiltonian in (2.11)
can then be written as in (2.2), 
$$
H_{\rm eff}={\Omega\over2}\, \vec n\cdot\vec\sigma\ ,
\eqno(3.4)
$$
in terms of a renormalized frequency
$$
\Omega=\omega +i\big[ {\cal K}(-\omega) - {\cal K}(\omega)\big]\ .
\eqno(3.5)
$$
As explained at the end of the previous section, a suitable acceleration
independent subtraction has been implicitly included in the definition
of the combination ${\cal K}(-\omega) - {\cal K}(\omega)$,
which otherwise would have been logarithmically divergent 
(compare with the result (2.23)).

In order to discuss the properties of the solutions of (2.11), (2.12)
it is convenient to express the density matrix $\rho$ in terms
of the Pauli matrices; recalling the normalization condition
${\rm Tr}[\rho]=1$, one has the standard expansion:
$$
\rho={1\over2}\Big(1 + \sum_{i=1}^3\, \rho_i\, \sigma_i\Big)\ .
\eqno(3.6)
$$
Then, the evolution equation (2.11) can be conveniently rewritten 
as a Schr\"odinger-like equation for the coherence (Bloch)
vector $|\rho(t)\rangle$ of components 
$\{\rho_1(t),\rho_2(t),\rho_3(t)\}$:[14]
$$
{\partial\over\partial t} |\rho(t)\rangle=-2\, {\cal H}\ |\rho(t)\rangle
+|\eta\rangle\ .
\eqno(3.7)
$$
The constant vector $|\eta\rangle$, with components
$\eta_i=-4B\, n_i$, $i=1,2,3$, comes from the imaginary
part of the Kossakowski matrix (3.2), while the
$3\times3$ matrix $\cal H$ includes contributions
both from $H_{\rm eff}$ and the real part of $a_{ij}$:
$$
{\cal H}_{ij}=\left[\matrix{a&b+\Omega_3&c-\Omega_2\cr
                                     b-\Omega_3&\alpha&\beta+\Omega_1\cr
                                     c+\Omega_2&\beta-\Omega_1&\gamma\cr}
									 \right]\ ,
\eqno(3.8)
$$
where $\Omega_i=(\Omega/2)\, n_i$, $i=1,2,3$, and
$$
\eqalign{
&a=2A+C\big(n_2^2+n_3^2\big)\cr
&\alpha=2A+C\big(n_1^2+n_3^2\big)\cr
&\gamma=2A+C\big(n_1^2+n_2^2\big)
}
\qquad\qquad
\eqalign{
&b=-C\, n_1 n_2\ ,\phantom{\big(n_1^2}\cr
&c=-C\, n_1 n_3\ ,\phantom{n_1^2}\cr
&\beta=-C\, n_2 n_3\ .\phantom{n_1^2}}
\eqno(3.9)
$$

The matrix $\cal H$ is non singular; indeed, its eigenvalues can be
explicitly determined: $\lambda_1=2A$, 
$\lambda_\pm=(2A+C)\pm i\Omega/2$. Further, their real parts are positive,
so that for large times $|\rho(t)\rangle$ reaches an equilibrium state
$|\rho_\infty\rangle$.[40]
This asymptotic state 
can be easily determined by inverting $\cal H$,
$$
|\rho_\infty\rangle={1\over2}\, {\cal H}^{-1} |\eta\rangle\ ,
\eqno(3.10)
$$
and turns out to be directed along the unit vector $\vec n$:
$$
|\rho_\infty\rangle= \bigg[{1-e^{\beta_U\omega}\over 1+e^{\beta_U\omega}}\bigg]\ |n\rangle\ .
\eqno(3.11)
$$
Inserting these components in the expansion (3.6), one finds that the
asymptotic density matrix $\rho_\infty$ is purely thermal, with a temperature
given by the Unruh temperature:
$$
\rho_\infty={e^{-\beta_U H_S}\over {\rm Tr}\big[e^{-\beta_U H_S}\big]}\ .
\eqno(3.12)
$$
Therefore, a two-level system which is uniformly accelerating 
in a vacuum scalar field
is driven to a thermal state with temperature $T_U$,
irrespectively of its initial state: this thermalization phenomenon
is the most obvious manifestation of the Unruh effect in the framework
of open system dynamics.

Nevertheless, further aspects of this phenomenon can be analyzed
by studying the behaviour of the solution of (3.7) for finite times,
that can be formally written as
$$
|\rho(t)\rangle={\cal M}(t)\, |\rho(0)\rangle +
\Big[1-{\cal M}(t)\Big]\, |\rho_\infty\rangle\ ,\qquad
{\cal M}(t)=e^{-2{\cal H} t}\ .
\eqno(3.13)
$$
The matrix ${\cal M}(t)$ is defined through the series expansion of the
exponential function and therefore seems to involve arbitrary powers of
$\cal H$. However, by definition, this $3\times3$ matrix obeys its cubic eigenvalue equation,
so that powers of $\cal H$ higher then two can always be reduced to combinations
of ${\cal H}^2$, $\cal H$ and ${\bf 1}$, the unit $3\times 3$ matrix.
Then, a systematic use of this substitution allows to write:
$$
{\cal M}(t)={4\over \Omega^2 +4C^2} \bigg\{ e^{-4At}\ \Lambda_1
+2\, e^{-2(2A+C)t}\bigg[ \Lambda_2 \cos\Omega t +
\Lambda_3\ {\sin\Omega t\over \Omega}\bigg]\bigg\}\ ,
\eqno(3.14)
$$
where the three constant $3\times3$ matrices $\Lambda_i$ are 
explicitly given by:
$$
\eqalign{
&\Lambda_1=\bigg[\big(2A+C\big)^2+{\Omega^2\over4}\bigg]\ {\bf 1}
-2\big(2A+C\big)\ {\cal H} + {\cal H}^2\ ,\cr
&\Lambda_2=-2A\big(A+C\big)\ {\bf 1} +\big(2A+C\big)\ {\cal H} 
-{1\over2}\, {\cal H}^2\ ,\cr
&\Lambda_3=2A\bigg[{\Omega^2\over4}- C\big(2A+C\big)\bigg]\ {\bf 1}
+\bigg[C\big(4A+C\big)-{\Omega^2\over4}\bigg]\ {\cal H}
-C\ {\cal H}^2\ .
}
\eqno(3.15)
$$
As expected, ${\cal M}(t)$ contains exponentially decaying factors
involving the real parts of the eigenvalues of $\cal H$, modulated
by oscillating terms in the effective frequency $\Omega$.
In other terms, an accelerating atom immersed in a vacuum scalar field is subjected to phenomena of decoherence and dissipation,
all regulated by the Planckian factors appearing in the
Kossakowski matrix (3.2), (3.3).

These non-unitary effects can be studied by analyzing
the time behaviour of suitable atom observables.
Indeed, any physical property  of the moving atom can be
represented by an hermitian matrix $\cal O$, that can be conveniently
decomposed as
$$
{\cal O}=\sum_{\mu=0}^3 {\cal O}_\mu\, \sigma_\mu\ .
\eqno(3.16)
$$
The time behaviour of its corresponding mean value is then
determined by that of the density matrix $\rho(t)$:
$$
\langle{\cal O}(t)\rangle={\rm Tr}\Big[{\cal O}\, \rho(t)\Big]=
{\cal O}_0 +\sum_{i=1}^3 {\cal O}_i\, \rho_i(t)\ .
\eqno(3.17)
$$
When the observable $\cal O$ represents itself an admissible 
atom state $\rho_{{\rm f}}$, the mean value (3.17) coincides
with the probability ${\cal P}_{{\rm i}\to {\rm f}}(t)$
that the evolved atom density matrix $\rho(t)$,
initially in $\rho(0)\equiv\rho_{{\rm i}}$,
be found in such a state at time $t$. Using (3.14), this probability
can be computed in general:
$$
\eqalign{
{\cal P}_{{\rm i}\to {\rm f}}(t)=&{1\over2}\bigg\{1-
\big(\vec\rho_{\rm f}\cdot\vec n\big)\, \Big(1-e^{-4At}\Big)\
\bigg[{1-e^{-\beta_U \omega}\over 1+e^{-\beta_U \omega}}\bigg]
+e^{-4At}\ \big(\vec\rho_{\rm i}\cdot\vec n\big)\, 
\big(\vec\rho_{\rm f}\cdot\vec n\big)\cr
&+e^{-2(2A+C)t}\ \Big(\Big[\big(\vec\rho_{\rm i}\cdot\vec\rho_{\rm f}\big)
-\big(\vec\rho_{\rm i}\cdot\vec n\big)\, 
\big(\vec\rho_{\rm f}\cdot\vec n\big)\Big]\ \cos\Omega t
-\vec n\cdot\big(\vec\rho_{\rm i}\times\vec\rho_{\rm f}\big)\
\sin\Omega t\Big)\bigg\}\ ,
}
\eqno(3.18)
$$
where, expanding the density matrices
$\rho_{\rm i}$, $\rho_{\rm f}$ as in (3.6), the notations
\hbox{$\big(\vec\rho_{\rm i}\cdot\vec\rho_{\rm f}\big)$}
and $\big(\vec\rho_{\rm i}\times\vec\rho_{\rm f}\big)$
(and similarly with $\vec n$)
represent scalar and vector products of their corresponding
coherence vectors.

When $\vec\rho_{\rm i}=-\vec n$ and $\vec\rho_{\rm f}=\vec n$,
the density matrices $\rho_{\rm i}$,
$\rho_{\rm f}$ represent the ground and excited states
of the system hamiltonian $H_S$ in (2.2) (see Appendix).
In this case, the expression in (3.18) simplifies,
$$
{\cal P}_{{\rm i}\to {\rm f}}(t)=
{1\over 1+e^{\beta_U \omega}}\ \Big(1- e^{-4At}\Big)\ ,
\eqno(3.19)
$$
giving the probability for a spontaneous transition of the atom from
the ground state to its excited state.
It is to this phenomenon of spontaneous excitation that one usually
refers when discussing the Unruh effect; 
indeed, (3.19) vanishes
as $\beta_U\to\infty$, {\it i.e.} for an atom at rest.

Although the behaviour of ${\cal P}_{{\rm i}\to {\rm f}}(t)$
in (3.18) and (3.19) is in principle experimentally observable
through the use of suitable interferometric devices,
in standard analysis of the Unruh effect one limits the
discussion to the spontaneous excitation rate 
$\mit\Gamma_{{\rm i}\to {\rm f}}$,
the probability per unit time of the transition
${\rm i}\to {\rm f}$, in the limit of an infinitely slow switching on
and off of the atom-field interaction.
In our formalism, its expression can be easily obtained 
by taking the time derivative of ${\cal P}_{{\rm i}\to {\rm f}}(t)$ 
at $t=\,0$; in the case of (3.19), one then finds
$$
{\mit\Gamma}_{{\rm i}\to {\rm f}}={\omega\over\pi}\,
{1\over e^{\beta_U \omega}-1}\ ,
\eqno(3.20)
$$
which is the expected result for an interaction hamiltonian of the form (2.3). 
One should nevertheless remark that the possibility of a non-vanishing 
$\mit\Gamma_{{\rm i}\to {\rm f}}$ is just one of the many manifestations
of the Unruh effect, that, as discussed above, actually involves 
phenomena of decoherence and dissipation;
in this respect, the open system approach to the description of
a moving atom dynamics appears to
be much more physically comprehensive 
than the most traditional treatments.

\vskip 2cm

{\bf 4. TWO ACCELERATING ATOMS AND ENTANGLEMENT\hfill\break}
\indent{\bf \phantom{4.} ENHANCEMENT}
\medskip

In the previous section we have seen that a uniformly accelerating
two-level atom immersed in a scalar field in its vacuum state
can be consistently described as an open system in weak interaction
with a heat bath: in the comoving frame, the atom is seen evolving
in time according to a master equation in Kossakowski-Lindblad
form, that through decoherence effects drives its state
towards a purely thermal equilibrium state, characterized
by the Unruh temperature. When the system that is subjected
to the uniform acceleration along the trajectory (2.1) is
formed by two, non-interacting two-level atoms, 
one thus expects similar mixing-enhancing phenomena to occur,
leading in particular to loss of the mutual quantum correlation 
(entanglement) that might have been present at the beginning.

However, even though not directly coupled, the external vacuum field
through which the two atoms move may provide an indirect 
interaction between them, and thus a means to entangle them.
Indeed, entanglement generation through the action of an external
heat bath has been shown to occur in certain circumstances;[18-23] 
it is therefore of physical interest to investigate the same issue
in the case of accelerating atoms.

We shall therefore start by considering a system 
composed by two, equal two-level atoms,
that start moving along the trajectory (2.1) at proper time $t=\,0$.
Being independent, without direct mutual interaction,
in the common comoving frame
their internal dynamics can again be taken 
to be described by the generic hamiltonian (2.2).
Then, the total two-system hamiltonian $H_S$ is now the sum of the
two terms:
$$
H_S=H_S^{(1)}+H_S^{(2)}\ ,
\qquad
H_S^{(1)}={\omega\over 2}\sum_{i=1}^3 n_i(\sigma_i\otimes\sigma_0)\ ,
\quad
H_S^{(2)}={\omega\over 2}\sum_{i=1}^3 n_i(\sigma_0\otimes \sigma_i)\ .
\eqno(4.1)
$$
Similarly, being immersed in the same field $\Phi_\mu$ and
within the weak-coupling hypothesis,
the atom-field interaction hamiltonian can be most simply assumed
to be the generalization of that in (2.3):
$$
H'=\sum_{\mu=0}^3\Big[\big(\sigma_\mu\otimes\sigma_0\big)
+\big(\sigma_0\otimes\sigma_\mu\big)\Big]\, \Phi_\mu\big(x(t)\big)\ .
\eqno(4.2)
$$
On the other hand, the field Hamiltonian $H_\Phi$ remains that of a collection
of free, independent scalar fields.

The derivation of the appropriate master equation describing the
dynamics of the two atoms in the comoving frame proceeds 
as in the case of a single moving atom, discussed in Section 2.
One starts from the Liouville-von Neumann equation (2.6) generating
the time evolution of the state $\rho_{\rm tot}(0)$
of the total system $\{ {\rm atoms}
+ {\rm external\ fields}\}$, and then traces over the fields degrees
of freedom, assuming a factorized initial state
$\rho_{\rm tot}(0)=\rho(0)\otimes |0\rangle\langle 0|$.
In the weak-coupling limit, the two-atom system density matrix $\rho(t)$
is seen evolving in time according to a quantum dynamical
semigroup of completely positive maps, generated by an equation
of Kossakowski-Lindblad form:
$$
{\partial\rho(t)\over \partial t}= -i \big[H_{\rm eff},\, \rho(t)\big]
 + {\cal L}[\rho(t)]\ .
\eqno(4.3)
$$
The unitary term depends on an effective hamiltonian
$H_{\rm eff}$ which is the sum of $H_S$ in (4.1) and suitable Lamb
contributions. In order to discuss them explicitly,
we proceed as in the single atom case and
adopt the simplifying condition (3.1), that results
in diagonal field correlations (2.14).
In this case, the effective hamiltonian consists
of the sum of three pieces: 
$H_{\rm eff}=H_{\rm eff}^{(1)}+H_{\rm eff}^{(2)}+H_{\rm eff}^{(12)}$.
The first two represent single system contributions;
they can be written exactly as in (4.1), with the frequency
$\omega$ replaced by the renormalized one $\Omega$
given in (3.5). The third term is a field-generated
direct two-atom coupling term:
$$
H_{\rm eff}^{(12)}=
i\sum_{i,j=1}^3\Big\{ \big[{\cal K}(\omega)+{\cal K}(-\omega)\big]\,\delta_{ij}
+\big[{\cal K}(0)-{\cal K}(\omega)-{\cal K}(-\omega)\big]\,n_i n_j\Big\}\
\sigma_i\otimes\sigma_j\ ,
\eqno(4.4)
$$
where ${\cal K}(\lambda)$ is the function introduced in (2.22).
As explained at the end of Section 2, 
a suitable acceleration independent subtraction has implicitly 
been included in the definition (4.4) in order to
make the contribution $H_{\rm eff}^{(12)}$ well defined.
Further, recall that ${\cal K}(\lambda)$ can be
split as in (2.23) into an acceleration dependent and an $a=\,0$ piece.
Since, as observed there, the acceleration dependent contribution
to ${\cal K}(\lambda)$ is odd in $\lambda$, one deduces that
$H_{\rm eff}^{(12)}$ does not
actually involves $a$: it is the same Lamb term that would have been
generated in the case of a two-atom system at rest. Being interested in 
acceleration induced effects, and in particular
in those related to entanglement creation, we shall not consider
any further this inertially generated term, nor the single
system contributions $H_{\rm eff}^{(1)}$, $H_{\rm eff}^{(2)}$
and move to analyze the effects produced by the dissipative
term ${\cal L}$ in (4.3).

Since the interaction of the two moving atoms with the
external scalar field is mediated by the same field operator
$\Phi(x)$, the Kossakowski matrix in ${\cal L}[\rho]$ involves
the same field correlation functions (2.14) discussed
in the case of a single atom system, and therefore
can be expressed in terms of the single atom
Kossakowski matrix $a_{ij}$ in (3.2), (3.3). Explicitly, one finds:[23]
$$
\eqalign{
{\cal L}[\rho]=
\sum_{i,j=1}^3\, a_{ij}\bigg(
&\bigg[(
\sigma_j\otimes\sigma_0)\,\rho\,(\sigma_i\otimes\sigma_0)
-{1\over2}\Big\{(\sigma_i\sigma_j\otimes\sigma_0)\,,\,\rho\Big\}
\bigg]\cr
+&\bigg[(
\sigma_0\otimes\sigma_j)\,\rho\,(\sigma_0\otimes\sigma_i)
           -{1\over2}\Big\{(\sigma_0\otimes\sigma_i\sigma_j)\,,\,\rho\Big\}
\bigg]\cr
+&\bigg[(\sigma_j\otimes\sigma_0)\,\rho\,(\sigma_0\otimes\sigma_i)
           -{1\over2}\Big\{(\sigma_i\otimes\sigma_j)\,,\,\rho\Big\}
\bigg]\cr
+&\bigg[(\sigma_0\otimes\sigma_j)\,\rho\,(\sigma_i\otimes\sigma_0)
           -{1\over2}\Big\{(\sigma_j\otimes\sigma_i)\,,\,\rho\Big\}
\bigg]
\bigg)\ .
}
\eqno(4.5)
$$

Describing the states of two two-level systems, the density matrix $\rho(t)$
is now a $4\times4$ matrix; in analogy with the decomposition (3.6) and recalling
the normalization condition ${\rm Tr}[\rho(t)]=1$, we find convenient
to decompose it as:
$$
\rho(t)={1\over4}\bigg[\sigma_0\otimes\sigma_0+
\sum_{i=1}^3 \rho_{0i}(t)\ \sigma_0\otimes\sigma_i
+ \sum_{i=1}^3 \rho_{i0}(t)\ \sigma_i\otimes\sigma_0
+ \sum_{i,j=1}^3 \rho_{ij}(t)\ \sigma_i\otimes\sigma_j\bigg]\ ,
\eqno(4.6)
$$
where the components $\rho_{0i}(t)$, $\rho_{i0}(t)$, $\rho_{ij}(t)$ are all real.
Substitution of this expansion in the master equation (4.3) allows deriving
the corresponding evolution equations for the above components of $\rho(t)$.
As explained above, we shall ignore the hamiltonian piece in (4.3)
since it can not give rise to acceleration induced entanglement. 
Further, we shall work in the regime of large acceleration,
{\it i.e.} in the limit of $\beta_U$ small; while
the conclusions concerning entanglement enhancement are actually
independent from this choice, it will make more
readable many explicit formulas.
In fact, with this simplifying assumption, the Kossakowski matrix
in (3.2) reduces to:%
\footnote{$^\dagger$}{Notice that the condition $B\leq A$, implicit in the original expressions
for $A$ and $B$ in (3.3), needs now to be formally imposed in order to
maintain the positivity of $a_{ij}$; it is physically justified
by the assumption of a small $\beta_U$.}
$$
a_{ij}=A\, \delta_{ij}-iB\, \epsilon_{ijk}\, n_k\ ,\qquad
A={1\over 2\pi\beta_U}\ ,\qquad B={\omega\over4\pi}\ .
\eqno(4.7)
$$

A straightforward but lengthy calculation allows then derive from
(4.3), and (4.5-7) the following result:
$$
\eqalignno{
{\partial\rho_{0i}(t)\over \partial t}=&-4A\,\rho_{0i}(t)+B(1+2\tau)\,n_i-
2B\sum_{k=1}^3 n_k\,\rho_{ik}(t)\ ,&(4.8a)\cr
{\partial\rho_{i0}(t)\over \partial t}=&-4A\,\rho_{i0}(t)+B(1+2\tau)\,n_i-
2B\sum_{k=1}^3 n_k\,\rho_{ki}(t)\ ,&(4.8b)\cr
{\partial\rho_{ij}(t)\over \partial t}=&-4A\big[2\,\rho_{ij}(t)
+\rho_{ji}(t)-\tau\,\delta_{ij}\big]+4B\big[n_i\,\rho_{0j}(t)+n_j\,\rho_{i0}(t)\big]\cr
&+2B\big[n_i\,\rho_{j0}(t)+n_j\,\rho_{0i}(t)\big]
-2B\, \delta_{ij}\,\sum_{k=1}^3 n_k\big[\rho_{k0}(t)+\rho_{0k}(t)\big]\ .&(4.8c)
}
$$
In these formulas, the quantity $\tau=\sum_{i=1}^3\rho_{ii}$ 
represents the trace of $\rho_{ij}$; 
it is a constant of motion, as easily seen by taking the trace
of both sides of $(4.8c)$. Further, the value of $\tau$ can not be
chosen arbitrarily, since it has to comply with the requirement
of positivity of the initial density matrix $\rho(0)$; indeed, using the
decomposition (4.6), one finds: $-3\leq\tau\leq 1$.

The system of first order differential equations in (4.8) naturally splits
into two independent sets, involving the symmetric,
$\rho_{(0i)}=\rho_{0i}+\rho_{i0}$, $\rho_{(ij)}=\rho_{ij}+\rho_{ji}$,
and antisymmetric,
$\rho_{[0i]}=\rho_{0i}-\rho_{i0}$, $\rho_{[ij]}=\rho_{ij}-\rho_{ji}$,
variables. Although both sets of equation can be exactly integrated,
the form of the explicit solutions looks cumbersome and is not
very inspiring. Nevertheless, by just examining the structure 
of the two sets of differential
equations, one can conclude that the antisymmetric variables
involve exponentially decaying factors, so that they vanish for
large times, while the symmetric variables approach in the same limit
a non-vanishing asymptotic value. As a consequence, the evolution
equations (4.8) admit an equilibrium state $\hat\rho$, whose explicit
form can be obtained by inverting the appropriate coefficient tensors
multiplying the variables $\rho_{0i}(t)$, $\rho_{i0}(t)$, $\rho_{ij}(t)$
in the r.h.s. of (4.8).  

The equilibrium density matrix $\hat\rho$
can be expanded as in (4.6): its components are given by:
$$
\eqalign{
&\hat\rho_{0i}=\hat\rho_{i0}={R\over 3+R^2}\big(\tau+3\big)\ n_i\ ,\cr
&\null\cr
&\hat\rho_{ij}={1\over 3+R^2}\Big[ \big(\tau-R^2\big)\ \delta_{ij}
+R^2\big(\tau+3\big)\ n_i\, n_j\Big]\ ,
}
\eqno(4.9)
$$
where $R=B/A$ is the ratio of the two constants appearing in the Kossakowski
matrix in (4.7), whose positivity implies: $0\leq R\leq1$. As expected,
the antisymmetric components $\hat\rho_{[0i]}$ and $\hat\rho_{[ij]}$ are zero,
while the only dependence on the initial state is through the
constant $\tau=\sum_i\rho_{ii}(0)$.%
\footnote{$^\dagger$}{One can check that the form (4.9) for the equilibrium
state remains unchanged even considering the more general evolution
generated by the Kossakowski matrix (3.2) instead of  
the simplified version in (4.7).
The considerations about entanglement production below are therefore
valid in general and not only in the case of large accelerations.}
It is remarkable that in general this equilibrium
state turns out to be entangled.

In the case of two, two-level systems, a measure of the entanglement content
of any state $\rho$ is provided by the concurrence ${\cal C}[\rho]$.[41-43] Indeed,
it has been shown that $\cal C$ is a monotonically increasing function
of the entanglement of formation;[44] its value ranges from zero, for separable
states, to one, for fully entangled states, like the Bell states.
In order to compute the concurrence of any $4\times4$ density matrix $\rho$
representing the state of two atoms, one starts from the auxiliary
matrix $\tilde\rho=(\sigma_2\otimes\sigma_2)\, \rho^T\, (\sigma_2\otimes\sigma_2)$,
where $T$ indicates transposition. Although not necessarily hermitian,
the matrix $\rho\tilde\rho$ has real, non-negative eigenvalues,
whose square roots $\lambda_\mu$, $\mu=1,2,3,4$, 
can be ordered in decreasing order: 
$\lambda_1\geq\lambda_2\geq\lambda_3\geq\lambda_4$. The concurrence of 
$\rho$ is then defined to be:
${\cal C}[\rho]={\rm max}\{\lambda_1-\lambda_2-\lambda_3-\lambda_4, 0\}$.

In the case of the asymptotic state $\hat\rho$ in (4.9), one finds that
the concurrence is indeed non-vanishing provided
$$
\tau< {5R^2-3\over 3-R^2}\ ,
\eqno(4.10)
$$
and that, in this case, its explicit expression is given by:
$$
{\cal C}[\hat\rho]={\big(3-R^2\big)\over2\big(3+R^2\big)}\,
\bigg[ {5R^2-3\over 3-R^2} -\tau\bigg]\ .
\eqno(4.11)
$$
The concurrence is therefore a linearly decreasing function of $\tau$,
starting from its maximum  ${\cal C}[\hat\rho]=1$ for $\tau=-3$ 
and reaching zero at $\tau=(5R^2-3)/(3-R^2)$;
notice that this ratio is an admissible value for $\tau$, since
it is always within the interval $[-1, 1]$ for the
allowed values of $R$. 

This result is remarkable, since it implies that the dynamics
in (4.8) can generate entanglement: one prepares
the two atoms in a separable state at $t=\,0$; then, 
provided the condition (4.10) is satisfied,
their long time equilibrium state will turn out to be entangled.
The simplest example of a separable state is provided
by the direct product of pure states:
$$
\rho(0)=\rho_n\otimes\rho_m\ , \qquad
\rho_n={1\over2}\Big(1 + \vec n\cdot\vec\sigma\Big)\ ,
\quad
\rho_m={1\over2}\Big(1 + \vec m\cdot\vec\sigma\Big)\ ,
\eqno(4.12)
$$
where $\vec n$ and $\vec m$ are two unit vectors.
In this case, one easily finds that $\tau=\vec n\cdot\vec m$,
so that, recalling (4.11), the asymptotic entanglement
is maximized when $\vec n$ and $\vec m$ are collinear
and pointing in opposite directions.
Explicitly, one finds:
$$
{\cal C}[\hat\rho]={2R^2\over 3+R^2}\ ,
\eqno(4.13)
$$
which reaches its maximum value, ${\cal C}[\hat\rho]=1/2$,
for $R=1$. 

The phenomenon of  entanglement production occurs also in cases when
the initial state $\rho(0)$
already has a non-vanishing concurrence.
As observed before, ${\cal C}[\hat\rho]$ reaches its maximum
when $\tau=-3$; in this case, the equilibrium state $\hat\rho$ coincides
with the totally entangled singlet state:
$$
\rho_-={1\over4}\bigg[\sigma_0\otimes\sigma_0
-\sum_{i=1}^3 \sigma_i\otimes\sigma_i\bigg]\ ,
\eqno(4.14)
$$
a fixed point of the dynamical equations (4.8),
as easily seen by direct inspection.
Then, let us consider the following initial state,
$$
\rho(0)=(1-\varepsilon)\rho_- +{\varepsilon\over4} \sigma_0\otimes\sigma_0\ ,
\eqno(4.15)
$$
which interpolates between $\rho_-$ and the totally mixed (separable) state;
for $\varepsilon< 2/3$, it is entangled, with 
${\cal C}[\rho(0)]=1-3\varepsilon/2$.
In passing from this initial state to its corresponding equilibrium
state $\hat\rho$ as $t\to\infty$, the corresponding increase
in concurrence, and thus of entanglement, can be easily computed:
$$
{\cal C}[\hat\rho]-{\cal C}[\rho(0)]={3R^2\varepsilon\over 3+R^2}\ ,
\eqno(4.16)
$$
which is indeed non vanishing. Further, notice that 
for the state (4.15), $\tau=-3(1-\varepsilon)$,
so that it can be taken to be very close 
to its lower limit $-3$; unfortunately, the entanglement production (4.16)
becomes vanishingly small as $\rho(0)$ approaches $\rho_-$.
In other terms, the maximally entangled state $\rho_-$
can never be asymptotically reached,
and thus the maximum of concurrence obtained, unless
one already starts with it at $t=\,0$.

\vskip 2cm

{\bf 5. DISCUSSION}
\medskip

Moving detectors, modelled as simple two-level atoms, 
immersed in external vacuum fields
and following an uniformly accelerating trajectory are
seen to posses a non-vanishing probability 
of spontaneous excitation, reproducing a
thermal spectrum: this phenomenon
is usually referred to as the Unruh effect.

It turns out that the dynamics of these accelerated atoms can be consistently
assimilated to that of subsystems in interaction with an external environment,
{\it i.e.} the so-called open quantum systems.
General techniques and results obtained in the analysis of the latter can then be
fruitfully applied to the study of the former, enabling the discussion of
physical aspects of the Unruh effect that might not be easily 
identified in the standard treatments. As shown in the previous sections,
the open system paradigm is particularly suitable for analyzing
on one hand issues connected to the appearance of decoherence
effects, and on the other hand questions related to 
the phenomenon of entanglement enhancement.

The starting point of our analysis has been the derivation of the appropriate
master equation that generates the time evolution of the states
of an accelerating system, in its comoving frame.
In the framework of a physically justified weak-coupling hypothesis, 
the system subdynamics
takes the form of a semigroup of completely positive maps,
transforming density matrices into density matrices,
while preserving their normalization and positivity.

In the case of a single two-level atom, the master equation
has been explicitly integrated; this has allowed discussing 
in detail the mixing-enhancing properties of the associated
finite time evolution, through the analysis of the behaviour
of appropriate atom observables.
In particular, one finds that
the moving atom is subjected to
dissipative effects, that asymptotically
drive its density matrix to an equilibrium state, with a purely
thermal spectrum.%
\footnote{$^\dagger$}{Note that this result is different
from (but complementary to) the ``thermalization theorem'' discussed
in the usual treatments of the Unruh effect: there,
it is the field ``vacuum'' state that appears to be purely thermal 
to an accelerating observer.}

When the accelerating system is composed by two, independent
atoms, its physical characterization naturally involves
the analysis of their mutual quantum correlations.
Because of the just mentioned decohering and mixing-enhancing phenomena,
one would be led to regard the Unruh effect
as counteracting entanglement enhancement.

Nevertheless, quite in general the presence of an external environment
(in the specific case of the external fields) can provide
indirect interaction between the two otherwise totally
decoupled two-level systems, thus a means to correlate them.
This picture has indeed been confirmed by the analysis 
of specific models; in particular,[23] entanglement can be
created by the action of an external bath through a purely
noisy mechanism during the memoryless, Markovian regime, when the
corresponding subdynamics is generated by equation
of the form (2.11), (2.12).

The master equation describing the dynamics of two accelerating atoms
discussed in Section 4 is precisely of the form identified in Ref.[23]
as generating initial entanglement. It is remarkable that 
the unavoidable decoherence that subsequently builds up
is not sufficient to counteract this quantum correlations enhancing effect.
The fate of the entanglement production by acceleration can in fact be
discussed by analyzing the entanglement content of
the final two-atom equilibrium state, through the evaluation
of its concurrence: quite in general, one finds an asymptotic
non-vanishing concurrence even for totally separable initial states.

Although obtained in an idealized situation, this result
offers new possibilities for an actual direct verification
of the Unruh effect. So far all efforts have been devoted
to the analysis of possible experimental settings
that could allow the measure of the tiny spontaneous excitation 
thermal rate induced in single accelerating systems.
Alternatively, using suitable devices,
one can instead try to detect the quantum correlation enhancement
that is generated when the accelerating system is formed
by two, initially un-entangled atoms; in view of the
high accuracy and sophistication reached by present quantum optics
experiments, this possibility might actually be realized
in the future.

As a concluding comment, let us remark that the presented open system
treatment of the Unruh effect is not limited to the analysis of
the simple setting of atoms interacting with free, scalar fields.
Extension to the case of higher spin fields is straightforward, while
the analysis of situations involving self-interacting fields 
would require the use algebraic quantum field theory;
this might not be such a formidable task as it looks, since the quantum
theory of open systems is amenable to a rigorous, algebraic
formulation. Finally, it is known that the Unruh effect 
has many similarities and analogies with the phenomenon of particle
creation in curved spacetimes. In this regard, we expect the open
system paradigm to be applicable also to those cases, possibly
providing new insight in the physical 
interpretation of the corresponding effects.

\vskip 2cm

\centerline{\bf ACKNOWLEDGEMENT}
\medskip

R.F. thanks Roberto Balbinot for many, useful discussions.

\vfill\eject

{\bf APPENDIX}
\medskip
 
Using the rigorous results of Refs.[16, 17],
we shall here present the derivation of the master
equation generating the reduced dynamics of a two-level
system in interaction with a generic external environment,
in the weak coupling limit. We shall generalize the
treatment of Ref.[45] and take the generic hamiltonian (2.2),
$$
H_S={\omega\over 2}\vec n\cdot\vec\sigma\ ,
\qquad |\vec n|=1\ ,
\eqno(A.1)
$$
as system hamiltonian; instead, the most
general term that is linear in both system
and environment variables,
$$
H'=\sum_{\mu=0}^3 \sigma_\mu\otimes B_\mu\ ,
\eqno(A.2)
$$
will be taken to represent the interaction hamiltonian.
The environmental dynamical variables $B_\mu$ are left unspecified
(they coincide with the field $\Phi_\mu$ in the case of 
an accelerating system), as the corresponding hamiltonian
$H_B$, that nevertheless is assumed to generate time translations:
$$
e^{iH_B t}\, B_\mu\, e^{-iH_B t}=B_\mu(t)\ .
\eqno(A.3)
$$

The time evolution of the density matrix $\rho_{\rm tot}$ representing the
state of the complete system is then generated by the total hamiltonian
$$
H=H_0+H'\ ,\qquad H_0=H_S\otimes{\bf 1}+{\bf 1}\otimes H_B\ ,
\eqno(A.4)
$$
through the standard unitary evolution, 
starting with an initial state taken
in factorized form: $\rho_{\rm tot}(0)=\rho(0)\otimes\rho_B$;
$\rho(0)$ is the $2\times2$ density matrix describing
the subsystem state, while $\rho_B$ is the analogous one
for the environment, assumed stationary: $[H_B,\rho_B]=\,0$
(in the case of moving atoms, $\rho_B\equiv|0\rangle\langle 0|$,
with $|0\rangle$ the Minkowski vacuum).

Correspondingly, the subdynamics describing the evolution
of the subsystem alone is obtained with a trace operation
over the environment degrees of freedom.
As briefly described in Section 2, and rigorously proven in Refs.[16, 17],
in the limit of weak coupling between subsystem and environment,
the reduced density matrix $\rho(t)={\rm Tr}[\rho_{\rm tot}(t)]$
is found obeying the following evolution equation:
$$
{\partial\rho(t)\over \partial t}= -i L_{H_S}[\rho(t)]
 + {\cal K}^\sharp[\rho(t)]\ ,\qquad L_{H_S}[\rho]\equiv\big[H_S,\,\rho\big]\ ,
\eqno(A.5)
$$
where
$$
{\cal K}^\sharp[\,\cdot\,]=-\lim_{T\to\infty}{1\over T}\int_0^T ds\
{\cal U}(-s)\ {\cal K}\ {\cal U}(s)\,[\, \cdot\,]\ ,\qquad
{\cal U}(s)=e^{-isL_{H_S}}\ ,
\eqno(A.6)
$$
and
$$
{\cal K}[\rho]=\int_0^\infty dt\ {\rm Tr}\Big( \big[e^{iH_0 t}\, H'\, e^{-iH_0 t},\big[H',\, 
\rho\otimes\rho_B\big]\big]\Big)\ .
\eqno(A.7)
$$
The ergodic mean over the system dynamics in $(A.6)$ is necessary in order to obtain
a completely positive subdynamics, and it is physically justified as an
averaging over the fast microscopic system oscillations; indeed, the weak coupling
limit procedure involves a suitable rescaling of the time variable, so that the equation
$(A.5)$ actually generates a ``coarse grained'' subdynamics with respect 
to the free system motion.
With the definitions $(A.1)$ and $(A.2)$, one can more explicitly write:
$$
\eqalign{
{\cal K}^\sharp[\rho]=\lim_{T\to\infty}{1\over T}\int_0^T\! ds
\int_0^\infty\!  dt&\, \Big[\sigma_\mu(t+s)\,\rho\,\sigma_\nu(s)\, \langle B_\nu\, B_\mu(t)\rangle
+\sigma_\nu(s)\,\rho\,\sigma_\mu(t+s)\ \langle B_\mu(t)\, B_\nu\rangle\cr
-&\sigma_\mu(t+s)\,\sigma_\nu(s)\,\rho\, \langle B_\mu(t)\, B_\nu\rangle
-\rho\,\sigma_\nu(s)\,\sigma_\mu(t+s)\, \langle B_\nu\, B_\mu(t)\rangle\Big]\ ,
}
\eqno(A.8)
$$
where 
$$
\langle B_\mu(t)\, B_\nu\rangle={\rm Tr}\big[ B_\mu(t)\, B_\nu\, \rho_B\big]\ ,
\eqno(A.9)
$$
are the environment correlations.

In order to proceed further, it is convenient to introduce the
two projectors operators
$$
P_\pm={1\over2}\Big(1\pm \vec n\cdot\vec\sigma\Big)\ ;
\eqno(A.10)
$$
they represent the density matrices of the two eigenstates of the
system hamiltonian $(A.1)$, with eigenvalues $\pm\omega/2$.
One can then use the auxiliary matrices
$\sigma_\mu^{(\xi)}$, $\xi=0, +, -$, explicitly defined by
$$
\sigma_\mu^{(0)}=P_+\, \sigma_\mu\, P_+ + P_-\, \sigma_\mu\, P_-\ ,\quad
\sigma_\mu^{(\pm)}=P_\pm\, \sigma_\mu\, P_\mp\ ,
\eqno(A.11)
$$
and represent the system free time evolution in terms of
the following spectral decomposition:
$$
\sigma_\mu(t)=e^{iH_S t}\, \sigma_\mu\, e^{-iH_S t}=\sum_{\xi=0,\pm}
e^{i\xi\omega t}\, \sigma_\mu^{(\xi)}\ .
\eqno(A.12)
$$
This allows performing explicitly the limit in $(A.8)$ and expressing
the result in terms of the following Fourier and Hilbert transform
of the environment correlations:
$$
\alpha_{\mu\nu}^{(\xi)}(\omega)=\int_{-\infty}^\infty dt\ e^{i\xi\omega t}\
\langle 0| B_\mu(t)\, B_\nu |0\rangle\ ,
\eqno(A.13)
$$
and
$$
\beta_{\mu\nu}^{(\xi)}(\omega)= \int_0^\infty dt\ e^{i\xi\omega t}\
\langle 0| B_\mu(t)\, B_\nu |0\rangle
-\int_0^\infty dt\ e^{-i\xi\omega t}\ \langle 0| B_\mu\, B_\nu(t) |0\rangle
\ ;
\eqno(A.14)
$$
notice that the first $4\times4$ matrix is hermitian,
$\big[\alpha_{\mu\nu}^{(\xi)}(\omega)]^\dagger=\alpha_{\mu\nu}^{(\xi)}(\omega)$,
while the second is antihermitian, 
$\big[\beta_{\mu\nu}^{(\xi)}(\omega)\big]^\dagger=-\beta_{\mu\nu}^{(\xi)}(\omega)$.
Explicitly, one finds:
$$
\eqalign{
{\cal K}^\sharp[\rho]={1\over2}\sum_{\xi=0,\pm}\ \sum_{\mu,\nu=0}^3\Big\{
&\alpha_{\mu\nu}^{(\xi)}\, \Big[ 2\sigma_\nu^{(-\xi)}\,\rho\,\sigma_\mu^{(\xi)}
-\sigma_\mu^{(\xi)}\sigma_\nu^{(-\xi)}\, \rho 
-\rho\, \sigma_\mu^{(\xi)}\sigma_\nu^{(-\xi)}\Big]\cr
&+\beta_{\mu\nu}^{(\xi)}\, \Big[ \rho,\, \sigma_\mu^{(\xi)}\sigma_\nu^{(-\xi)}\Big]\Big\}\ .
}
\eqno(A.15)
$$
This expression can be further simplified by expanding the auxiliary
matrices $\sigma_\mu^{(\xi)}$ in terms of Pauli matrices:
$$
\sigma_0^{(\xi)}=\delta_{\xi 0}\, \sigma_0\ ,\qquad
\sigma_i^{(\xi)}=\sum_{j=1}^3 \psi_{ij}^{(\xi)}\, \sigma_j\ ,
\eqno(A.16)
$$
with
$$
\psi_{ij}^{(0)}=n_i\, n_j\ ,\qquad 
\psi_{ij}^{(\pm)}={1\over 2}\big(\delta_{ij} - n_i\, n_j\pm i\epsilon_{ijk} n_k\big)\ .
\eqno(A.17)
$$
The entire master equation in $(A.5)$ can then be rewritten 
in standard Kossakowski-Lindblad form:
$$
{\partial\rho(t)\over \partial t}= -i \big[H_{\rm eff},\, \rho(t)\big]
 + {\cal L}[\rho(t)]\ ,
\eqno(A.18)
$$
where
$$\eqalignno{
&{\cal L}[\rho]={1\over2} \sum_{i,j=1}^3 a_{ij}\big[2\, \sigma_j\rho\,\sigma_i 
-\sigma_i\sigma_j\, \rho -\rho\,\sigma_i\sigma_j\big]\ ,&(A.19a)\cr
&H_{\rm eff}={1\over2}\sum_{i=1}^3\big[\omega\, n_i +b_i\big]\ ,&(A.19b)\cr
}
$$
while the Kossakowski matrix $a_{ij}$ and hamiltonian vector $b_i$ can be
expressed as:
$$
\eqalignno{
&a_{ij}=\sum_{\xi=0,\pm}\ \sum_{k,l=1}^3\ \alpha_{kl}^{(\xi)}\, \psi_{ki}^{(\xi)}\,
\psi_{lj}^{(-\xi)}\ ,&(A.20a)\cr
&b_i=i\sum_{j=1}^3 \Big[\alpha_{0j}^{(0)}-\alpha_{j0}^{(0)}
-\beta_{0j}^{(0)}-\beta_{j0}^{(0)}\Big] n_j\, n_i\cr
&\hskip 2cm +\sum_{j,k,l,m=1}^3 \epsilon_{ijk}\ \bigg[ \sum_{\xi=0,\pm}\ \,\beta_{lm}^{(\xi)}\
\psi_{lj}^{(\xi)}\,\psi_{mk}^{(-\xi)}\bigg]\ . &(A.20b)
}
$$
Since $\big[\psi_{ij}^{(\xi)}\big]^*=\psi_{ij}^{(-\xi)}$, 
the matrix $a_{ij}$ in $(A.20a)$ results manifestly hermitian and positive, 
being the combination of Fourier transform of
correlation functions.[14] 
Further, by choosing for the environment variables $B_\mu$ the fields $\Phi_\mu$,
the master equation $(A.18-20)$ reduces to that discussed in the text.

Notice that in general the dissipative piece in $(A.19a)$ is a function of
nine, real parameters, the independent entries of the matrix $a_{ij}$.%
\footnote{$^\dagger$}{These are not actually all independent: 
since the matrix $a_{ij}$ is positive,
they need to satisfy certain inequalities; see Ref.[28] for details.}
However, because of the structure given in $(A.20a)$, the Kossakowski matrix 
obtained through a weak coupling procedure appears to depend on a lesser
number of free parameters. This fact has been already noticed in Ref.[45],
where it is further observed that the most general master equation obtainable
in the weak coupling limit 
appears to coincide with the old Bloch equation, describing
the dissipative motion of a spin in a constant magnetic field.

In the light of the previous derivation, this conclusion looks however 
too restrictive:
by starting with the most general system hamiltonian $(A.1)$, instead
of the one with $\vec n=(0,0,1)$ as adopted in Ref.[45], one is able to obtain a master
equation in the form $(A.18-20)$, certainly more general than
the Bloch equation. This observation might have interesting
applications in the study of specific open system models.

\vskip 3cm

\centerline{\bf REFERENCES}
\vskip 1cm

\item{1.} W.G. Unruh, Phys. Rev. D {\bf 14} (1976) 870
\smallskip
\item{2.} P.C.W. Davies, J. Phys. A {\bf 8} (1975) 609
\smallskip
\item{3.} N.D. Birell and P.C.W Davies, {\it Quantum Fields in Curved
Space}, (Cambridge Univ. Press, Cambridge, 1982)
\smallskip
\item{4.} W.G. Unruh and R.M. Wald, Phys. Rev. D {\bf 29} (1984) 1047
\smallskip
\item{5.} S. Takagi, Prog. Theor. Phys. Suppl. {\bf 88} (1986) 1
\smallskip
\item{6.} R.M. Wald, {\it Quantum Field Theory in Curved Spacetime
and Black Hole Thermodynamics}, (Chicago University Press, Chicago, 1994)
\smallskip
\item{7.} G.L. Sewell, Ann. of Phys. {\bf 141} (1982) 201
\smallskip
\item{8.} R. Haag, {\it Local Quantum Physics: Fields, Particles, Algebras},
(Springer, Berlin, 1996)
\smallskip
\item{9.} For a recent review, see: H.C. Rosu, Frenet-Serret vacuum radiation,
detection proposals and related topics, {\tt hep-th/0301128}
\smallskip
\item{10.} B.S. DeWitt, in {\it General Relativity, An Einstein
Centenary Survey}, S.W. Hawking and W. Israel, eds.,
(Cambridge University Press, Cambridge, 1979)
\smallskip
\item{11.} E.B. Davies, {\it Quantum Theory of Open Systems}, (Academic Press,
New York, 1976)
\smallskip
\item{12.} V. Gorini, A. Frigerio, M. Verri, A. Kossakowski and
E.C.G. Surdarshan, Rep. Math. Phys. {\bf 13} (1978) 149 
\smallskip
\item{13.} H. Spohn, Rev. Mod. Phys. {\bf 53} (1980) 569
\smallskip
\item{14.} R. Alicki and K. Lendi, {\it Quantum Dynamical Semigroups and 
Applications}, Lect. Notes Phys. {\bf 286}, (Springer-Verlag, Berlin, 1987)
\smallskip
\item{15.} H.-P. Breuer and F. Petruccione, {\it The Theory of Open
Quantum Systems} (Oxford University Press, Oxford, 2002)
\smallskip
\item{16.} E.B. Davies, Comm. Math. Phys. {\bf 39} (1974) 91
\smallskip
\item{17.} E.B. Davies, Math. Ann. {\bf 219} (1976) 147
\smallskip
\item{18.} D. Braun, Phys. Rev. Lett. {\bf 89} (2002) 277901
\smallskip
\item{19.} M.S. Kim, J. Lee, D. Ahn and P.L. Knight,
Phys. Rev. A {\bf 65} (2002) 040101(R)
\smallskip
\item{20.} S. Schneider and G.J. Milburn, Phys. Rev. A {\bf 65} (2002) 042107
\smallskip
\item{21.} A.M. Basharov, J. Exp. Theor. Phys. {\bf 94} (2002) 1070
\smallskip
\item{22.} L. Jakobczyk, J. Phys. A {\bf 35} (2002) 6383
\smallskip
\item{23.} F. Benatti, R. Floreanini and M. Piani, Phys. Rev. Lett.
{\bf 91} (2003) 070402
\smallskip
\item{24.} A. Peres and D.R. Terno, Rev. Mod. Phys. {\bf 76} (2004) 93
\smallskip
\item{25.} F. Strocchi, Relativistic quantum mechanics and field theory,
{\tt hep-th/0401143}
\smallskip
\item{26.} S.S. Schweber, {\it An Introduction to Relativistic Quantum
Field Theory}, (Harper and Row, New York, 1961)
\smallskip
\item{27.} D. Ahn, H. Lee and S.W. Hwang, Phys. Rev. A {\bf 67} (2003) 032309
\smallskip
\item{28.} V. Gorini, A. Kossakowski and
E.C.G. Surdarshan, J. Math. Phys. {\bf 17} (1976) 821 
\smallskip
\item{29.} G. Lindblad, Commun. Math. Phys. {\bf 48} (1976) 119
\smallskip
\item{30.} F. Benatti and R. Floreanini,
Banach Center Publications, {\bf 43} (1998) 71
\smallskip
\item{31.}  F. Benatti and R. Floreanini, Mod. Phys. Lett. A {\bf 12} (1997) 465
\smallskip
\item{32.} F. Benatti and R. Floreanini, Nucl. Phys. {\bf B511} (1998) 550
\smallskip
\item{33.} F. Benatti, R. Floreanini and R. Romano, J. Phys. A
{\bf 35} (2002) L551
\smallskip
\item{34.} F. Benatti, R. Floreanini and M. Piani, Phys. Rev. A 
{\bf 67} (2003) 042110
\smallskip
\item{35.} R. Dumcke and H. Spohn, Z. Physik {\bf B34} (1979) 419
\smallskip
\item{36.} K.-P. Marzlin and J. Audretsch, Phys. Rev. D {\bf 57} (1998) 1045
\smallskip
\item{37.} J. Audretsch and R. M\"uller, Phys. Rev. A {\bf 52} (1995) 629
\smallskip
\item{38.} H.A. Bethe, Phys. Rev. {\bf 72} (1947) 339
\smallskip
\item{39.} P.W. Milonni, {\it The Quantum Vacuum: An Introduction
to Quantum Electrodynamics}, (Academic Press, San Diego, 1994)
\smallskip
\item{40.} K. Lendi, J. Phys. A {\bf 20} (1987) 13
\smallskip
\item{41.} S. Hill and W.K. Wootters, Phys. Rev. Lett. {\bf 78} (1997) 5022
\smallskip
\item{42.} W.K. Wootters, Phys. Rev. Lett. {\bf 80} (1998) 2245
\smallskip
\item{43.} W.K. Wootters, Quantum Inf. Comp. {\bf 1} (2001) 27
\smallskip
\item{44.} C.H. Bennett, D.P. DiVincenzo, J. Smolin and W.K. Wootters, Phys. Rev. A
{\bf 54} (1996) 3824
\smallskip
\item{45.} K. Lendi, $N$-level sysyems and application to spectroscopy, 
in Ref.[14]

\bye